# FULL-DUPLEX COMMUNICATIONS

## Performance in Ultra-Dense Small-Cell Wireless Networks

Animesh Yadav, Georgios I. Tsiropoulos, and Octavia A. Dobre

*Abstract*— Theoretically, full-duplex (FD) communications can double the spectral-efficiency (SE) of a wireless link if the problem of self-interference (SI) is completely eliminated. Recent developments towards SI cancellation techniques have allowed to realize the FD communications on low-power transceivers, such as small-cell (SC) base stations. Consequently, the FD technology is being considered as a key enabler of 5G and beyond networks. In the context of 5G, FD communications have been initially investigated in a single SC and then into multiple SC environments. Due to FD operations, a single SC faces residual SI and intra-cell co-channel interference (CCI), whereas multiple SCs face additional inter-cell CCI, which grows with the number of neighboring cells. The surge of interference in the multi-cell environment poses the question of the feasibility of FD communications. In this article, we first review the FD communications in single and multiple SC environments and then provide the state-of-the-art for the CCI mitigation techniques, as well as FD feasibility studies in a multi-cell environment. Further, through numerical simulations, the SE performance gain of the FD communications in ultra-dense massive multiple-input multiple-output enabled millimeter wave SCs is presented. Finally, potential open research challenges of multi-cell FD communications are highlighted.

## Introduction

Greater device affordability has been driving increased smartphone adoption, which results in a tremendous increase in the subscriptions to cellular and broadband services. Besides, billions of Internet-of-Things (IoT) devices are expected to rely on cellular wireless networks. Based on a recent Ericsson's mobility report published in June 2017, it is anticipated that by the end of 2022 there will be 29 billion connected devices, which translates into an immense volume of wireless traffic. To meet this growing demand for wireless access, the research community strains to include in the emerging fifth generation (5G) and beyond cellular networks the most advanced wireless communication technologies. Among them, the full-duplex (FD) technology has lately gained attention due to its potential of theoretically doubling the spectral efficiency (SE) when compared to half-duplex (HD) communications. In addition, the FD communications bring several other potential benefits, as they avoid the hidden terminal problem, enhance network secrecy, improve sensing in cognitive radio networks and reduce the end-to-end packet delay/latency. Providing low-latency is one of the key features of the 5G New Radio (NR) interface. The FD communications can be considered as a realizable technology to offer low-latency for NR.

### FD Radio Characteristics

FD communications facilitate the simultaneous transmission and reception of data over the same frequency band. However, due to superimposition of leaked high-power transmit and low-power received signals at the FD transceiver, FD suffers from self-interference (SI), which hampers its operation. Recently, several works have reported important developments in SI cancelation (SIC) techniques. Essentially, a combination of passive and active techniques has been adopted for SIC. Although the SI is usually compensated for in the digital domain, this cannot compensate for the noise resulting from the analog circuitry of the transmitter. Accordingly, SIC relies on both analog and digital techniques, also known as active SIC. Additionally, passive SIC techniques can be employed, which are applied at the antenna level, e.g., physical separation between transmit and receive antennas. The combined use of the available SIC techniques attenuates the SI power to the thermal noise level. Experimental results have shown that this level of attenuation is sufficient for FD operation and identified low-power transceivers, such as small-cell (SC) base stations (SBSs), as potential deployment scenario.

### From Single-Cell to Multi-Cells

Cellular operators worldwide are increasingly introducing small-size cells, such as micro-, pico- and femto-cells, within the coverage area of macro-cells to improve the capacity and coverage of networks. In the context of 5G and beyond wireless networks, researchers advocate the deployment of even smaller size cells and their densification, which basically means a low user-SBS ratio. Furthermore, SCs are inexpensive and easy to deploy. In addition, the dense deployment of SCs improves the spatial frequency reuse, facilitates user offloading from macro BS to SBS, and reduces the transmit power.

Since the deployment of the FD transceiver on the low-power SBSs is possible, several works investigated FD communications in SCs. Earlier studies have been performed to verify the capacity gain offered by the FD communications in a single SC. The main challenge the single SC faces is co-channel interference (CCI) from the uplink (UL) to the downlink (DL) transmission besides the residual SI (RSI) after SIC. Moving from single-cell to realistic multi-cell scenario introduces a huge surge of CCIs including both intra- and inter-cell CCI. For comparison purpose, it is important to know how the number of CCI links increases when moving from the HD to the FD communications. Figure 1 graphically depicts the CCIs arisen in a multi-cell conventional HD communications, where the DL and UL transmissions occur on orthogonal frequency bands. On the other hand, Figure 2 (a) depicts the number of CCI links when each SBS operates in the FD mode, i.e., both DL and UL transmissions occur on the same frequency band simultaneously. Recently, a few works have appeared that investigate the feasibility of FD communications in the multi-cell environment over HD communications. The results urge for novel design of intelligent radio resource allocation schemes and schedulers in order to extract the FD benefits.

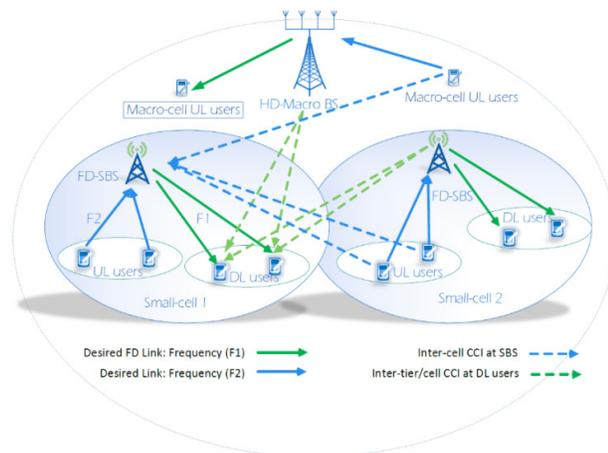

**Figure 1** Co-channel interference links in two SCs. Each SC has one HD SBS, two DL and UL users. The DL and UL communications are performed on orthogonal frequency F1 and F2, respectively.

In the rest of this article, firstly, the FD communications in both single and multiple SC networks are presented. Secondly, the state-of-the-art interference mitigation schemes in multiple SC networks are introduced. Additionally, the feasibility of FD communications with 5G key enabling technologies, such as massive MIMO (mMIMO) and millimeter wave (mmWave), is discussed in the context of improving the channel characteristics, reducing the interference introduced in the network, and addressing the problem of SI. Thirdly, quantitative SE performance gain obtained due to FD communication in ultra-dense SC networks is provided. Fourthly, potential future research directions are discussed, and finally, conclusions are drawn.

## FD in Small-Cell Networks

### The Single-Cell Case

In a single-cell, three deployment scenarios are common due to their applicability in 5G and beyond networks: i) multiuser MIMO systems with FD BSs serving HD/FD users; ii) MIMO FD relays; iii) FD ad-hoc links, such as device-to-device (D2D) communications. In the first scenario, the FD SBS communicates on the UL and DL channels with the HD users, as shown in Figure 2 (a) [1]. In the case of FD users, the BS communicates with the same user on the UL and DL channels simultaneously [2]. As such, the FD system has to cope not only with the RSI, but also with the intra-cell CCI from the UL user towards the DL user. In [2], multiple antennas are used at both SBS and users and optimal precoding and resource allocation schemes are proposed for achieving the improved SE due to FD communications. Furthermore, the degree-of-freedom (DoF) due to the multiple antennas is exploited for SIC as well [3].

In the second scenario, fixed FD relays assist the distant users to communicate with the BS within a cell [4]. As a consequence, there is a gain in terms of both time and frequency resources, since the end-to-end communication is realized within a single time slot and over a single channel.

In the third scenario of ad-hoc communication, especially D2D links are considered in the FD-based cell [4]. A challenging scenario involves the direct communication of devices employing D2D links in the licensed band, operating in a distributed fashion. Simultaneously, the same channel may be used by the wireless cellular communications. Since the distributed D2D communications are not controlled by the BS, severe issues arise regarding the interference management, which in turn require advanced spectrum sensing processes and user cooperation.

As the next generation networks are anticipated to bank on key enabling technologies such as mMIMO, feasibility studies of the FD communications within this context are underway [5]. In mMIMO, the order of hundreds antennas is employed when compared with the current systems. In a single SC, the FD SBS employing mMIMO benefits the system as follows: i) provides higher throughput and reliability; ii) allows simpler signal processing techniques, such as zero-forcing (ZF) and maximum ratio combining (MRC); and iii) reduces the SI power significantly with the increase in the number of transmit antennas [5].

### Moving Towards the Multi-Cell Environment

Moving from the single- to the multi-cell scenario is not only a matter of increasing coverage and capacity. The FD communications in the multi-cell environment are envisioned for both wireless access and backhaul networks. In wireless access, within each SC, the DL and UL users simultaneously communicate with the FD SBS on the same frequency spectrum. On the other hand, in-band backhaul network allows

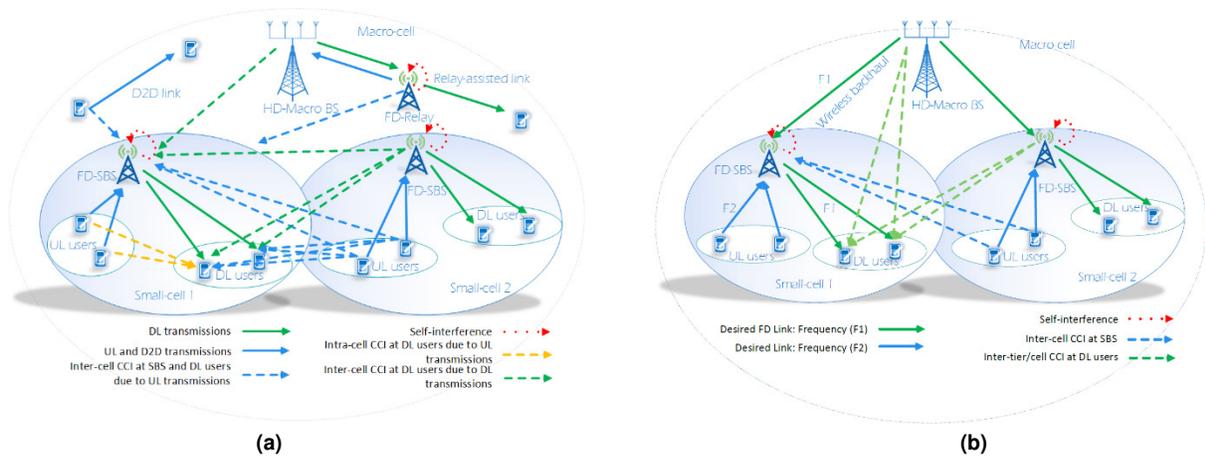

**Figure 2 (a)** Co-channel interference links in two SCs. Each SC has one FD SBS, a few DL and UL users, as well as ad-hoc D2D users. **(b)** Co-channel interference links arise due to in-band FD wireless backhaul communications at the SBS.

the FD SBSs to communicate simultaneously over the access and backhaul links in the same frequency spectrum. In both cases, the interference environment in multi-cell FD networks is significantly more complex. Besides the RSI and intra-cell CCI, inter-cell CCI occurs in wireless access, as shown in Figure 2 (a). While there is no intra-cell CCI from the UL to DL users, the backhaul network suffers from inter-cell CCI as shown in Figure 2 (b). Additionally, a larger number of neighboring SCs introduces increased aggregated inter-cell CCI. Taking into consideration the complicated interference environment generated by the application of FD in multi-cell networks, it is not clear whether the overall network SE improves or not when compared to that of the HD systems.

The studies that offer a clearer perception of the FD performance in the interference limited multi-cell network environment generally use mathematical frameworks based on stochastic geometry [6], [7], [8] and convex-optimization [9], [10], [11], [12], [13], [14]. Stochastic geometry is a mathematical tool used in modeling and analysing the performance of large dense wireless networks with random topologies. In [6], a hybrid HD/FD two-tier multi-cell heterogeneous network is considered. Results demonstrate that by allowing the different tiers to operate in different duplex modes enhances the overall network SE. In [7], FD SBSs are considered, where in each SC, an HD UL user communicates with another HD DL user through the SBS. The study advocates that under a moderate RSI power level, and higher cell density where inter-cell CCI becomes strong, equipping both SBSs and DL users with multiple antennas is beneficial in mitigating the CCI. The multiple antennas are essentially used for signal-to-interference-noise ratio (SINR) maximization, via MRC technique; this brings larger SE gains than using them for supressing the inter-cell CCI via the ZF method. An investigation on whether to deploy FD tranceivers on macro- or micro-cells networks in an interference-limited multi-user scenario is conducted in [8] under the assumption that SI is sufficiently cancelled. Results show that without proper countermeasures for CCI management, the FD communication is infeasible on the macro-cell network. However, if the micro BSs are apart by a certain critical distance, then the FD mode enhances the SE of micro-cell networks.

On the other hand, the framework based on convex-optimization involves the design of advanced radio resource management and interference mitigation techniques [9]- [14]. Furthermore, we consider combining the FD communications with mMIMO and mmWave technologies, which help in mitigating both SI, and intra- and inter-cell CCIs.

## Interference Management in Multi-Cells

In this section, we discuss the state-of-the-art interference mitigation techniques and technologies used in multi-cell environments.

### Interference Mitigation Techniques

The techniques that are used for mitigating the intra- and inter-cell CCIs can be broadly classified as: i) interference alignment (IA); ii) interference-aware user pairing/clustering; iii) traffic-aware user scheduling; and iv) beamforming design.

In the first class, an intra- and inter-cell CCI mitigation strategy based on IA is employed [9]. On the UL channel, transmit beamformers are designed for aligning the UL user-to-SBS and UL user-to-DL user CCIs. On the other hand, on the DL channel, each SBS designs the transmit beamformers for aligning the DL multiuser interference (MUI), SBS-to-DL user and SBS-to-SBS CCIs. The achievable sum DoF using the IA technique in the FD multi-cell environment increases with the number of cells, users, and SBS antennas.

In the second class, the SBSs opportunistically select the FD/HD modes based on the intra-cell CCI among the users. It has been shown in [9] that intra-cell CCI is dominant when compared with the inter-cell CCI. Therefore, when intra-cell

CCI is low (high), the SBS selects FD (HD) mode for transmission [10], [11]. However, the aggregated inter-cell CCI is still significant, and hence, requires adequate mitigation techniques to deal with it for achieving larger SE gain. The interference levels strongly depend on the user locations, traffic demand, etc. To deal with both intra- and inter-cell CCI, each SC first selects or pairs the appropriate users to maximize its SE, then the scheduler coordinates with other SCs to allocate the power levels of selected users such that the aggregate network SE is maximized.

In the third class, intra- and inter-cell CCI can be reduced by scheduling the user based on their data rate requirements [12]. In the previously discussed approaches, the interference mitigation schemes have been developed by maximizing the SE of the networks. In reality, every user has a different number of backlogged bits queued-up in a data buffer for transmission, and hence, minimizing the queue length of the data buffer is an appropriate metric. Moreover, this approach limits the resource allocations beyond the number of backlogged bits without explicit rate constraints. Consequently, the scheduler schedules all the users in the network in such a way that their sum queue length is minimized. After the end of each scheduling instant, a user is not scheduled if it receives zero rate. Moreover, there can be some SCs where only one or zero users are scheduled. The scheduled user could be either on the DL or UL channel, i.e., HD mode is selected for that particular SC. The SBS does not participate in the network if no users are scheduled in that SC. In other words, this approach implicitly selects the hybrid HD/FD mode for the SBSs based on the traffic and interference levels.

Lastly, in the fourth class, interference management can be performed without explicitly avoiding the intra-cell CCI as discussed previously [9]- [11]. The interference mitigation primarily depends on the multiple antennas employed by both SBSs and users. The scheduler collects all the required channel state information (CSI) and then maximizes the SE of the network by employing optimal transmit and receiver beamformers [13].

Note that the scheduler can operate either in centralized or decentralized manner. A centralized scheduler requires a large amount of global CSI acquisition of every link involved before it makes a final decision. On the other hand, a decentralized scheduler makes a final decision with a fewer information exchange, and hence, it has practical applicability.

*Interference Mitigation Technologies*

The interference in FD communications can be further suppressed by using the mMIMO and mmWave technologies.

*Massive MIMO:* A potential alternative strategy to address the huge surge of interference introduced in FD multi-cell networks is to exploit the very large DoF offered by mMIMO [14]. In the multi-cell scenario, mMIMO can be realized at FD SBSs likewise in a single SC case [5]. The increased number of SBS antennas helps to form highly directional beams. Additionally, at the same time, it proportionally scales down the transmit powers of both SBSs and users while maintaining the quality-of-service. The scaled down powers are proportional to the reciprocal of the number of SBS antennas; consequently, the intra- and inter-cell CCI levels are reduced. For perfect CSI and infinite number of antennas, the interference-limited multi-cell FD system asymptotically doubles the SE over the HD system.

*Millimeter Wave:* mmWave is the promising technology that enhances the SE of wireless networks since it operates over an abundant frequency spectrum; thus, it can achieve high-speed wireless communications, as demonstrated by the recent 802.11ad Wi-Fi at 60 GHz. In mmWave bands, the signals suffer large path loss, shadowing, and blockage; accordingly, highly directional beams are essential in achieving high SINR at the user. Hence, combining the mmWave with FD and mMIMO technologies reduces the RSI significantly as a highly directional beam has a weak line-of-sight (LOS) component towards its own receiver antennas. Consequently, the UL users can transmit with lower power; this results in a reduction of the intra-cell CCI, which is dominant when compared with inter-cell CCI [10], [11]. Furthermore, FD SC-based networks can benefit from the employment of mmWave communications, since signals have reduced travel range and strength. The short range of mmWave signals, when compared to the microwave one, significantly decreases the interference due to FD between neighboring cells, as well as the MUI within the SC range.

## Numerical Results

In this section, we present the throughput performance of a network of multiple SCs, which employs mMIMO and mmWave technologies, through numerical simulations. The network throughput is obtained by minimizing the total data queue buffer length of a multiple SC network. A two-tier heterogeneous network in an urban outdoor environment is considered, as shown in Figure 3. In tier one, there is a circular macro-cell region with an HD macro BS positioned at the origin. In the second tier, multiple FD SCs are positioned according to the Poisson point process with density $\lambda_s$ (km$^{-2}$) within the macro-cell region. Each SC has uniformly positioned two HD DL and two UL user within its coverage region. Further, it is assume that each DL and UL user has a backlog of six bits in its data buffer. We are interested in reducing the total amount of backlog bits of UL and DL users of the network. Hence, we employ a centralized scheduler, which uses an optimization algorithm as presented in [12].

The considered SC-based network can operate in both licensed and unlicensed frequency bands from the sub-6 GHz and mmWave spectra. In simulations, we consider a frequency band from the 28 GHz spectrum, which helps in realizing the mMIMO configuration easily on the SBSs. Furthermore, the channels between user-to-SBS, SBS-to-user, and user-to-user are modeled based on the description provided in [15]. Each

SBS has two radio-frequency (RF) chains and employs hybrid precoding, whereas the UL users have a single RF chain and employ analog precoding for transmission. Each SBS, in the DL channel, transmits the superimposed data to its users. While decoding, each DL user considers intra- and inter-cell users data as interference. On the other hand, for multiuser detection in the UL channel, the SBS performs minimum mean square error successive interference cancellation for the same cell users and considers the received signal from other cells as interference. The important parameters and their values used in simulations are summarized in Table 1.

| Parameters | Value |
|---|---|
| Number of antennas at SBS | Transmit: 64 and Receive: 1 |
| Number of antennas at user | Transmit: 16 and Receive: 1 |
| Cell radius | MBS: 500 m, SBS: 100 m |
| Maximum transmit powers | SBS: 24 dBm, User: 23 dBm |
| Carrier frequency and system bandwidth | 28 GHz and 100 MHz |
| Thermal noise density and RSI power | -174 dBm/Hz, -110 dB |
| Path loss exponents | LOS: 2.1 and NLOS: 3.4 |
| Channel model | mmWave channel [15] |

**Table 1** Simulation parameters.

We study the various modes selected by the SBSs and the total queue deviation of the network with varying SC deployment densities. The SBS can select either OFF (no transmission), HD or FD mode of operation. Figure 4 shows the percentage of these modes of operation selected by the SBSs with varying SC densities in a macro-cell range. It can be observed that a larger SBS number selects the FD mode when the SC density is small, whereas more SBSs select the OFF and HD modes of operation when the density increases. The reason for this trend is the level of CCI, including both inter- and inter-cell, received by the SBSs. In low-density regime, the CCI levels are low, and the SBSs prefer the FD mode. On the other hand, in the high-density regime, the majority of the SBSs receive high levels of CCI, and thus, avoid the FD mode. Furthermore, the SBSs cease their operations when the levels of CCI are very high.

Figure 5 compares the performances of FD-SBS and HD-SBS. In simulations, we assume that the HD-SBSs serve only their DL user. It can be observed that the FD mode outperforms the HD mode. Furthermore, the increase in the SC density results in increased inter-cell CCI, and hence, the total queue deviation increases.

## Challenges and Broader Perspectives
The previous section investigated the potential of FD in multi-cell wireless networks, as well as the relevant benefits in terms of total network SE. Although we could conclude that FD works in multi-cell networks, there still exist demanding challenges to address while achieving a sufficient level of compatibility with existing LTE systems.

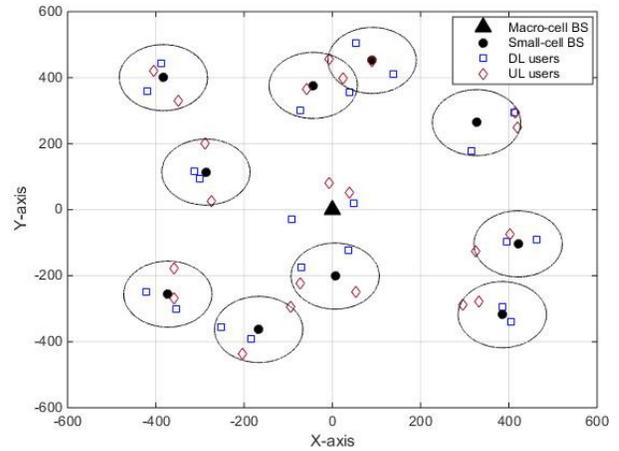

**Figure 3** SCs network with an HD MBS at the centre and ten FD SBSs. Each SC serves two HD DL and two HD UL users.

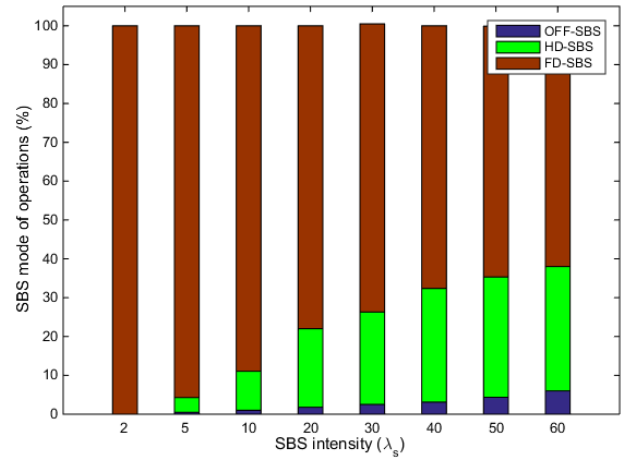

**Figure 4** Percentage of the SBSs which select among the OFF, HD and FD modes versus the SBS density within a macro-cell region.

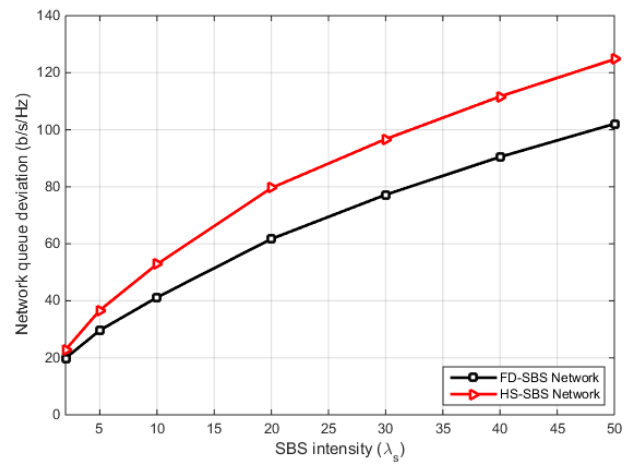

**Figure 5** Network queue deviation versus SBS density within a macro-cell region.

Recently, non-orthogonal multiple access (NOMA) has been considered as an alternative to the traditional orthogonal multiple access. This method not only increases the SE, but also improves user fairness and reduces the latency at the medium access layer. NOMA enables the multiplexing of users simultaneously utilizing the same time-frequency resources either in power- or code-domain at the expense of the receiver complexity. It is anticipated that integrating NOMA with FD further enhances the system capabilities, such as increasing the number of users per time-frequency resource. Despite foreseen benefits, their integration is not straightforward. For example, due to NOMA, additional multiuser interference is introduced over the same channel besides the SI and CCI due to the FD communications. Furthermore, successive interference cancelation should be performed at the receivers for multiuser detection, whose complexity increases with the number of users. To this end, investigating the achievable improvement in SE of FD-NOMA over HD-NOMA or pure HD systems in multi-cell networks, which are equipped with mMIMO and operate in mmWave spectrum band is an interesting and unexplored research direction.

Acquiring perfect knowledge of CSI from all the nodes involved in a large-scale multi-cell network is rare due to limited feedback or fast channel variations. Moreover, the set of components employed at the transceivers are prone to hardware impairments, such as phase noise, nonlinearities in the power amplifiers, I/Q imbalance and analog-to-digital converter impairments. Since the FD multi-cell networks are interference-limited, these imperfections affect the overall SE gain of a network. Furthermore, they are even more detrimental to FD-NOMA multi-cell networks. Hence, considering them in the network designs to study the feasibility of FD/FD-NOMA in multi-cell network represents a viable research direction.

Drones or unmanned aerial vehicles have been actively considered as aerial SBSs for 5G and beyond networks. In particular, aerial SBSs are supposed to provide wireless services in dangerous and disastrous regions, and in those affected by high blockage. However, the drones are hover-time constrained, and hence, the effective use of hover duration is vital for the aerial SBSs. So far, the HD transceivers have been considered for an aerial SBS. Investigating the FD transceivers is an interesting research direction as the FD SBS can improve the SE for the same flying time.

Another type of wireless network, which is dedicated for establishing the communication between vehicles, is the vehicular ad-hoc network (VANETs). Low latency, better traffic coordination, safety and high reliability are the major concerns for VANET. Also, VANET is expected to provide a cooperative awareness, i.e., the availability to individuate the moving vehicles in the surroundings for improved traffic safety. The FD transceivers are a good choice for improving the latency, as they enable simultaneous transmission and reception. Furthermore, FD transceivers improve the cooperative awareness by finding a larger number of neighboring vehicles, especially in heavy traffic conditions. Despite the FD communications benefits, self-interference and co-channel interference limit the performance concerning the latency, reliability, and safety. Hence, exploring the prospect of FD communications in VANET is a meaningful direction of research.

## Conclusion

This article provides a brief overview of the state-of-the-art developments in FD communications both in single- and multi-cell environments. Various intra- and inter-cell CCI mitigation techniques are described. It has been shown that with intelligent interference mitigation schemes, FD communications are feasible and beneficial in an SC networks. Furthermore, due to the inherent characteristics of mMIMO and mmWave technologies, it has been recognized that they not only enhance the SE but also mitigate the CCI erupted in a multi-cell environment. Numerical simulations were conducted for an interference-limited mMIMO-enabled mmWave multi-SC environment. Results have shown the SE gain offered by FD over HD communications. Further, in an ultra-dense deployment, less SBSs prefer to operate in FD mode. Additionally, major open research challenges in FD multi-cell environments are highlighted.

## Biographies

ANIMESH YADAV (animeshy@mun.ca) is a Research Associate at Memorial University, Canada. Previously, he worked as a postdoctoral fellow and research scientist at UQAM, Canada, and CWC at University of Oulu, Finland, respectively. During 2003-07, he worked as a software specialist at iGate Global Solution, India and Elektrobit Oy, Finland. He received his Ph.D. degree from University of Oulu, Finland. He is recipient of the best paper awards at IEEE WiMOB-2016 and IWCMC-2017. His research interests include enabling technologies for future wireless networks and green communications.

GEORGIOS I. TSIROPOULOS (gitsirop@mail.ntua.gr) received the Diploma, M.Sc., and Ph.D. degrees in electrical and computer engineering from National Technical University of Athens, Greece, in 2005, 2009, and 2010, respectively. He served with the Ministry of Transport and Communications of the Hellenic Republic (2006-2009) and the UHC, Greece, (2011-2015) as an ICT consultant. Since 2015, he has been an ICT project manager at the Hellenic Electricity Distribution Network Operator S.A. His current research activities focus on cooperative communications and resource optimization in 5G systems.

OCTAVIA A. DOBRE (odobre@mun.ca) is a Professor and Research Chair at Memorial University, Canada. In 2013 she was a Visiting Professor at Massachusetts Institute of Technology, USA, and University of Brest, France. Previously, she was with New Jersey Institute of Technology, USA, and Polytechnic Institute of Technology, Romania. She was the recipient of a Royal Society scholarship and a Fulbright fellowship. Her research interests include enabling technologies for 5G, cognitive radio systems, blind signal identification and parameter estimation techniques, as well as optical and underwater communications. She has authored around 200 referred journal and conference papers in these areas. Dr. Dobre is the EiC of the IEEE Communications Letters. She has served as editor for various prestigious journals, and technical co-chair of different international conferences, such as IEEE ICC and Globecom. She is a member-at-large of the Administrative Committee of the IEEE Instrumentation and Measurement Society and served as chair and co-chair of different technical committees. Dr. Dobre is a Fellow of the Engineering Institute of Canada.